\documentstyle[aps,prb,epsf,twocolumn,floats]{revtex}
\begin{document}
\draft
\twocolumn[\hsize\textwidth\columnwidth\hsize\csname @twocolumnfalse\endcsname

\title{Logarithmic corrections to finite size spectrum of SU(N) 
symmetric quantum chains}
\author{Kingshuk Majumdar$^{(1)\!}$ and Manash Mukherjee$^{(2)\!}$}
\address{$^{(1)\!}$ Department of Physics, Berea College, Berea, KY 40404,\\ 
$^{(2)\!}$ Department of Physics, Miami University, Oxford, OH 45056.}
\maketitle
\begin{abstract}
{We consider $SU(N)$ symmetric one dimensional quantum chains at finite 
temperature. For such systems the correlation lengths, ground state 
energy, and excited state energies are investigated in the framework 
of conformal field theory. The possibility of different types of 
excited states are discussed. Logarithmic corrections to the ground 
state energy and different types of excited states in the presence 
of a marginal opeartor, are calculated. Known results for $SU(2)$ and 
$SU(4)$ symmetric systems follow from our general formula.}
\end{abstract}
\bigskip
]
\narrowtext 
One of the fundamental models of solid state physics is the Heisenberg 
model of insulating magnets. In the one-dimensional case (``spin 
chains''), the spin-$1/2$ Heisenberg models have been studied 
extensively: most of our understanding of their quantum critical 
behavior is based on the Bethe-Ansatz solution for the ground state 
and excitation spectrum,\cite{bethe,luther} mapping to the Sine-Gordon 
theory,\cite{yang} non-abelian bosonization,\cite{aff2} and mapping to 
the sigma model.\cite{haldane} Although spin-$1/2$ Heisenberg chains 
are $SU(2)$ symmetric systems, fruitful generalizations have been 
accomplished in two different directions: (a) enlarging the 
representation of $SU(2)$ group to study quantum chains with higher 
spins, and (b) introducing higher symmetry groups such as $SU(N)$.

Here we consider generalizations of the type (b) and investigate how 
higher symmetry affects the ground state properties 
[Eqs.~(\ref{gen})-(\ref{xip})] and finite size spectrum of quantum 
``spin'' chains.  Earlier studies of Affleck\cite{aff1} show that any 
one dimensional system with $SU(N)$ symmetry is critical, and at the 
very low energy scale these models are equivalent to $(N - 1)$ free 
massless bosons. This system of free bosons - when viewed in the 
framework of two dimensional conformal field theory, are the primary 
fields of the $SU(N)_{k=1}$ WZNW model. Adopting this model we give 
an explicit derivation of logarithmic corrections to finite size 
spectrum of $SU(N)$ symmetric quantum chains. Logarithmic shifts 
in excited states energy levels have been theoretically observed 
for $SU(2)$ and $SU(4)$ symmetric systems away from the $T=0$ quantum 
critical point.~\cite{aff5,itoi,aff4,mila,azar,ueda} Known results for 
$N=2$ and $N=4$ follow from our general formulas [obtained in 
Eqs.~(\ref{bnfinal})-(\ref{bfinal}) and the paragraphs below 
Eq.~(\ref{bnb})].

An one-dimensional $SU(N)$ symmetric quantum  
chain of length $L$ is described by the Hamiltonian~\cite{bill}
\begin{equation}
H = \sum_{j=1}^n \sum_{A=1}^{N^2-1}S^A_j S^A_{j+1}.
\label{ham}
\end{equation}
\noindent 
Here $n=L/a_0$ is the total number of discrete points ($a_0$ being 
the lattice spacing) and $S^A_j$ are the $(N^2-1)$  
generators of the $SU(N)$ Lie algebra at 
each lattice site $j$. For convenience, the interaction strength and 
the lattice spacing $a_0$ have been set equal to one in 
Eq.~(\ref{ham}). At each site $j$, the generators $S^A_j$ can be 
represented by $N$ ``flavors'' of fermions, $\psi_{aj}$  [$a=1,...,N$]
\begin{equation}
S^A_j = \sum_{a,b=1}^N\psi^{a\dagger}_j (T^A)^b_a\psi_{bj} - {I/N},
\label{sa} 
\end{equation} 
where $I$ is the identity operator, and $T^A$'s are a complete set of 
$(N^2-1)$ traceless normalized matrices so that ${\rm Tr}[ T^AT^B ] = 
(1/2) \delta^{AB}$.  Eq.~(\ref{sa}) satisfies the constraint that at 
each site the total number of fermions is conserved, i.e. 
$\sum^N_{a=1} \psi^{a \dagger}_j \psi_{ja} = 1$.

For such a fermionic system the theory can be bosonized using 
non-abelian bosonization at low temperatures. In the continuum limit, 
the bosonized Hamiltonian\cite{aff2} ($H_{\rm eff}$) can then be written in terms 
of the Kac-Moody currents, 
\begin{equation} 
H_{\rm eff} \approx v_s 
\sum_{A=1}^{N^2-1}\int\!dx \left[ J^A_LJ^A_L + J^A_RJ^A_R + 
2J^A_LJ^A_R\right],
\label{bosham} 
\end{equation}
where the normal ordered Kac-Moody currents for the left (with Fermi momentum 
$k_F<0$) and the right moving (with $k_F>0$) fermions are defined as,
\begin{eqnarray} 
J^A_L &=& :\psi^{a \dagger}_L (T^A)^b_a \psi_{Lb}:, \nonumber \\ 
J^A_R &=& :\psi^{a \dagger}_R (T^A)^b_a\psi_{Rb}:.  
\end{eqnarray}
At zero temperature ($T=0$) the interaction term, 
$\sum_{A=1}^{N^2-1}J^A_LJ^A_R$, in Eq.~(\ref{bosham}) renormalizes to 
zero, and the sum of first two terms in the Hamiltonian that are 
quadratic in left and right moving currents, corresponds to the 
$SU(N)_{k=1}$ WZNW model. The fundamental unitary $N \times N$ matrix 
field $g$ of the WZNW model is given by,
\begin{equation}
g^a_b =  {\rm (const)}:\psi^{a\dagger}_L \psi_{Rb}:.  
\end{equation}
The field $g$ transforms as the fundamental representation of 
${ SU(N)}_{L} \times { SU(N)}_R$ which describes the exact symmetry
of the Hamiltonian in Eq.~(\ref{bosham}) at zero temperature.
It is known that this fermionic theory is equivalent to a 
theory of $(N-1)$ free massless bosons at the criticality with 
velocities $v_s$; they correspond to $(N-1)$ excitation modes of 
the $SU(N)$ symmetric quantum chain that oscillates at different values of 
$k_F$.~\cite{aff1}  Furthermore, these oscillating modes are primary fields of the 
$SU(N)_{k=1}$ WZNW model and their scaling dimensions ($\Delta_p$) 
can be obtained from~\cite{kz}
\begin{equation}
\Delta_p = \frac {2C_p}{C_{\rm adj}+1}.
\end{equation}
For $SU(N)$, $C_p\;(p=1,~2, \cdots, N-1)$ is the eigenvalue of the 
Casimir operator in the $p$-th fundamental representation [having a 
Young tableau with $p$ boxes in a single column], and $C_{\rm adj}=N$ 
is the eigenvalue of the Casimir operator in the adjoint representation 
[having a 2-column Young tableau with $(N-1)$ boxes in the first 
column and 1 box in the second column].  In a highest weight 
($\Lambda$) representation of $SU(N)$, the corresponding Casimir 
eigenvalue is given by\cite{fuchs}
\begin{equation}
C_{\Lambda} = \frac {(\theta, \theta)}{2}\left[m\left(N - \frac m{N}\right) 
+ \sum_{i=1}^{r_o} (b_i)^{2} - \sum_{i=1}^{c_o} (a_i)^{2}\right],
\label{casm}
\end{equation}
where $\theta$ is the highest weight corresponding to the adjoint 
representation and {\bf is normalized to 1}, $m$ is the total number 
of boxes in the Young tableau with $r_o$ rows of length $b_1, b_2, 
\cdots, b_{r_o}$ and $c_o$ columns of length $a_1, a_2, \cdots, 
a_{c_o}$. Using this formula we find $C_p=p(N-p)(N+1)/2N$ and $C_{\rm 
adj}=N$. Hence, the scaling dimensions ($\Delta_p$) of primary 
fields of the $SU(N)_{k=1}$ WZNW model are given by
\begin{equation}
\Delta_p = \frac {2C_p}{C_{\rm adj}+1} = \left[\frac {p(N-p)}{N} \right].
\label{gen}
\end{equation}
For example, in the case of $SU(4)$ the three oscillating components 
have scaling dimensions (3/4,~1,~3/4) [for $p=1,~2,~3$].  For $SU(N)$, the 
mode (dominant) that oscillates at $k_F=2\pi/N$ has a scaling 
dimension $(1-1/N)$ [$p=1$ or $(N-1)$ in this case].

Finite-size corrections to Heisenberg chain with $SU(2)$ and $SU(4)$ 
symmetry has been studied using conformal field 
theory.~\cite{{aff5},{cardy}} The relevance of studying finite size 
chains are twofold.  One can not only compare the theoretical results 
with numerical simulations and experiments which are limited to finite 
size of the system but also can study the finite temperature behavior 
of the system by identifying the finite size in the imaginary time 
direction, that corresponds to finite temperature. To obtain the 
finite-size corrections of a 1D chain of length $L$ and with periodic 
boundary conditions we first introduce a conformal mapping from the infinite 
plane (with coordinate $z$) to the cylinder (with coordinate $w$) via 
$w= (L/2\pi)\ln z$. Identifying the length to be inverse of the 
temperature [$L=v_s/T$] the finite temperature results\cite{cardy} of the ground 
state energy $E_0$ can be generalized to the $SU(N)$ symmetric system,
\begin{equation}
E_0 (T)= E_0 (0) - \frac {\pi T(N-1)}{6v_s}.
\label{groundstate}
\end{equation}
Here $E_0(0)$ refers to the ground state energy at zero temperature.  
The thermodynamic quantities like specific heat and entropy can now be 
obtained by taking the appropriate derivatives with respect to the 
temperature.

Other quantities of interest are the finite temperature corrections to 
the correlation lengths ($\xi$) of the different modes.  These inverse 
of the correlation lengths, $\xi^{-1}$ are signature of energy gaps, 
$(E_n-E_0)$ between the ground state and the lowest lying excited 
states ($E_n$) that are created by finite temperature of the system.  
Using the general formula for the scaling dimension 
[Eq.~(\ref{gen})], we obtain $ \xi_p^{-1}$ of the $p$-th staggered 
mode,
\begin{eqnarray}
\xi^{-1}_p & \equiv & E_n^p-E_0, \nonumber \\
&=& \left(\frac {2\pi T}{v_s}\right) \Delta_p = 
\left(\frac {2\pi T}{v_s}\right)\left[\frac {p(N-p)}{N} \right].
\label{xip}
\end{eqnarray}

The temperature dependence of the correlation lengths is in fact 
modified by logarithmic corrections in the presence of marginal 
operators in the theory.~\cite{cardy} The generic form of the 
Hamiltonian at the critical point containing a marginal operator $\phi 
(x,t)$ is
\begin{equation}
H = H^* + g_o\int\! dx \;\phi, 
\end{equation}
where $g_o$ is the coupling constant and $H^*$ is the Hamiltonian at 
the fixed point. In our case [Eq.~(\ref{bosham})], the normalized marginally 
irrelevant operator is
\begin{equation} 
\phi = -D \sum_{A=1}^{N^2-1} J_L^A J_R^A.
\label{mar}
\end{equation} 
For such a marginally irrelevant operator, the Hamiltonian at the 
critical point ($T =0$) becomes equal to the fixed point Hamiltonian.  
In Eq.~(\ref{mar}), $D$ is the normalization constant to be determined from 
the two-point correlator of $\phi$.  The operator product expansion 
(OPE) of the $J_{L(R)}^A$ with any normalized Kac-Moody primary field 
$\chi$ is given by~\cite{cft}
\begin{eqnarray} 
J_{L}^A (z) \chi (z') &=& \frac {J^A_{0,L}}{2\pi i (z-z')}\chi (z') + \cdots,
\label{OPE1}\\
J_{R}^A ({\bar z}) \chi ({\bar z}') &=& \frac {J^A_{0,R}}{2\pi i
({\bar {z}-\bar {z}'})}\chi ({\bar z}') + \cdots,
\label{OPE2}  
\end{eqnarray}
where the operators $J^A_{0,L}$ and $J^A_{0,R}$ are the generators of 
the global ${ SU(N)}_{L} \times { SU(N)}_R$ transformations, 
and satisfy the characteristic equations, $J^A_{0,L}|\chi 
(z')\rangle = -T_{L}^A |\chi (z')\rangle $ and $J^A_{0,R}|\chi 
({\bar z}')\rangle = |\chi ({\bar z}')\rangle T_{R}^A$.  Note that 
$\sum_A(J^A_{0,L})^2$ and $\sum_A(J^A_{0,R})^2$ are the Casimir 
operators of ${\rm SU(N)}_L$ and $SU(N)_R$ groups, respectively. The 
two-point correlators of the left currents (for $k=1$) are,
\begin{eqnarray}  
\langle J_{L}^A (z) J_{L}^B (z') \rangle &=& -\frac { {\rm Tr} [T^A_{L} 
T^B_{L}]}{4\pi^2 (z-z')^2}= -\frac {\delta^{AB}}{8\pi^2 (z-z')^2}, \\
\langle J_{R}^A ({\bar z}) J_{R}^B ({\bar z}') \rangle &=& 
\frac {\delta^{AB}}{8\pi^2 ({\bar z}-{\bar z}')^2}.
\end{eqnarray}  
Using these results we explicitly calculate: 
\begin{equation} 
\langle \phi (z,{\bar z}) \phi (z',{\bar z}') \rangle 
=\left( \frac {D}{8\pi^2}\right)^2 \frac {(N^2-1)}{(z-z')^2({\bar z} -
{\bar z}')^2},
\end{equation}
and then compare it to the standard conformal field theory result i.e.  
$\langle \phi (z,{\bar z}) \phi (z',{\bar z}') \rangle = 
|z-z'|^{-2}|{\bar z}-{\bar z}'|^{-2}$ to obtain the value of the 
constant,
\begin{equation}
D = \frac {8 \pi^{2}}{\sqrt{N^2-1}}.
\end{equation}
Thus the normalized irrelevant marginal operator is given by,
\begin{equation}
\phi (z,{\bar z}) = -\frac {8 \pi^{2}}{\sqrt{N^2-1}} 
\sum_{A=1}^{N^2-1} J_L^A(z) J_R^A({\bar z}).
\label{marginal}
\end{equation}

Perturbation to the normalized excited state ($\phi_n$) energies due 
to the marginal operator can now be calculated~\cite{cardy} from
\begin{equation}
\delta (E_n - E_0) = g_{o} \int\! d x \;\langle \phi_n | 
\phi | \phi_n \rangle,
\label{energydiff}
\end{equation}
where $\phi$ and $\phi_n$ are Virasoro primary fields generated by 
applying Fourier modes of $J_L^A$ and $J_R^A$ on Kac-Moody primary 
fields. For large length 
(equivalently, small temperature), we may replace the coupling $g_0$ 
by its renormalization group improved value (upto the $\log - \log$ 
term),\cite{nomura}
\begin{eqnarray}
g_0(T)&=& \left( \frac 1{\pi b \ln (T_0/T)}\right)  \nonumber \\
& \times & \left[1 - \frac {1}{2\ln (T_0/T)}\ln[\ln (T_0/T)] \right]. 
\label{gvalue}
\end{eqnarray}
Here $T_0$ is the model dependent parameter of the system and the 
coefficient $b$ is defined via the following 3-point correlator,
\begin{equation}
\langle \phi (z_1,{\bar z}_1) \phi (z_2, {\bar z}_2) 
\phi (z_3,{\bar z}_3)\rangle = -b/|z_{12}|^{2}|z_{23}|^{2}|z_{13}|^{2}.
\label{beqn}
\end{equation}
Substituting this in Eq.~(\ref{energydiff}) we obtain,
\begin{eqnarray}
\delta (E_n &-& E_0) = \left(\frac {2 \pi T}{v_s\ln (T_0/T)}\right) 
\nonumber \\
&\times & \left(\frac {2b_n}{b}\right)\left[1- \frac 1{2\ln(T_0/T)}
\ln[\ln(T_0/T)] \right].
\label{energycorr}
\end{eqnarray}
The coefficient $b_n$ is again defined through the 3-point correlator,
\begin{eqnarray}
& &\langle \phi_n (z_1,{\bar z}_1) \phi (z_2, {\bar z}_2) 
\phi_n (z_3,{\bar z}_3)\rangle \nonumber \\
&=& - b_n/|z_{12}|^{2}|z_{23}|^{2}|z_{13}|^{2x_n-2}.
\label{bn}
\end{eqnarray}
Here $x_n$ is the scaling dimension of the Virasoro primary field 
$\phi_n$.  Substituting Eq.~(\ref{marginal}) in Eq.~(\ref{bn}) and 
using the OPE's as in Eqs.~(\ref{OPE1}) and (\ref{OPE2}) it follows 
that $b_n$ is directly proportional to the sum of the product of the 
eigenvalues of the generators $J^A_{0,L}$ and $J^A_{0,R}$,
\begin{equation}
b_n = -\frac {2}{\sqrt{N^2-1}}\sum_{A=1}^{N^2-1}T^A_LT^A_R.
\label{newbn} 
\end{equation}  
To evaluate $\sum_A T^A_LT^A_R$, we observe that the full 
symmetry, $SU(N)_{L}\times SU(N)_{R}$, of the quantum chain at the critical 
point is broken by the presence of the marginal operator, 
$\phi(z,{\bar z})$. Only the diagonal $SU(N)\subset SU(N)_{L}\times 
SU(N)_{R}$ is an exact symmetry of the quantum chain. Under this 
subgroup, the representation $V_{L}\otimes V_{R}$ of 
$SU(N)_{L}\times SU(N)_{R}$ decomposes into direct sum of various 
irreducible subrepresentations. If an excited state 
($|\phi_n\rangle$) belongs to a highest weight subrepresentation, 
$V\subset V_{L}\otimes V_{R}$ and $C$ is the corresponding Casimir invariant of 
the diagonal $SU(N)$ in $V$, then we have\cite{etin}
\begin{equation}
\sum_{A=1}^{N^2-1} T^A_LT^A_R = \frac {1}{2}[C-C_L-C_R],
\label{iden}
\end{equation}
where $C_{L}$ and $C_{R}$ are the Casimir invariants of $SU(N)_L$ and 
$SU(N)_R$ in the highest weight representations $V_{L}$ and $V_{R}$, respectively. 
Therefore, using Eqs.~(\ref{newbn}) and (\ref{iden}) we find
\begin{equation}
b_n = -\frac 1{\sqrt{N^2-1}}\left[C-C_L-C_R \right].
\label{bnfinal}
\end{equation}
The above formula may also be used to evaluate the renormalization
group coefficient, $b$ [Eq.~(\ref{beqn})]: since $\phi (z,{\bar z})$
is a Virasoro primary field of conformal dimensions (1,1), we set
$\phi_{n} (z,{\bar z}) = \phi (z,{\bar z})$ and $x_{n} = 2$ in
Eq.~(\ref{bn}), and hence, $b_{n} = b$.  This can be seen as follows. 
The Virasoro primary fields $J^A_{L}$ and $J^A_{R}$ of conformal
dimensions $(1,0)$ and $(0,1)$ transform as the adjoint
representations, $V_{L}^{\rm adj}$ and $V_{R}^{\rm adj}$ of $SU(N)$, and since
$V_{R}^{\rm adj}$ is conjugate to $V_{L}^{\rm adj}$, the direct sum
decomposition of $V_{L}^{\rm adj}\otimes V_{R}^{\rm adj}$ under the diagonal
$SU(N)\subset SU(N)_{L}\times SU(N)_{R}$ must contain a unique
singlet. Hence, the Virasoro primary field $\phi (z,{\bar z})$ in
Eq.~(\ref{marginal}) transforms as this singlet representation and we have
$C=0, C_L=C_R = N$ in Eq.~(\ref{bnfinal}).  This implies
\begin{equation}
b = \frac {2N}{\sqrt{N^2-1}}.
\label{bfinal}
\end{equation}
For example in the case of Heisenberg spin-chain with SU(2) symmetry, 
$b=4/\sqrt{3}$ and for the spin-orbital model with SU(4), 
$b=8/\sqrt{15}$. Our result for SU(4) is new.  
Together with Eq.~(\ref{gvalue}), the constant $b$ also 
determines the correction of $O(g_{o}^3)$ in the ground state energy,
\begin{equation}
E_0 (T)-E_0 (0)=-\left(\frac {\pi T}{6v_s}\right) \left[(N-1) + 2\pi^3b g_{o}^3\right].
\label{ground}
\end{equation}

To determine the logarithmic shifts in the excited states energy 
levels, we need the ratio $2b_n/b$ in Eq.~(\ref{energycorr}). From 
Eqs.~(\ref{bnfinal}) and (\ref{bfinal}) we get,
\begin{equation}
\left(\frac {2b_n}{b}\right) = -\frac 1{N}\left[C-C_L-C_R \right].
\label{bnb}
\end{equation}

To evaluate $b_n$ (and hence $2b_n/b$) we must know the excited 
states. In $SU(N)$ invariant quantum chains, the low lying excited 
states ($|\phi_n\rangle$) correspond to $(N-1)$ primary fields of the 
$SU(N)_{k=1}$ WZNW model with the scaling dimensions, $\Delta_{p}$.  
These fields transform as $(q,{\bar q})$ representations of 
$SU(N)_{L}\times SU(N)_{R}$, where
\begin{eqnarray}
q & = & \frac {N(N-1)\cdots(N-p+1)}{p!}\nonumber 
\end{eqnarray}
is the dimension of the $p$-th fundamental representation of $SU(N)$ 
for $p=1,~2, \cdots, N-1$. For instance, the lowest excited states 
correspond to the fundamental primary field $g$ 
with the scaling dimensions, $\Delta_{1}=(1 - 1/N)$.  This 
field transforms under the $(N,{\bar N})$ representation which 
decomposes under the diagonal $SU(N)$ into the adjoint and singlet 
representations. For the excited states, ${\rm Tr}\;[\,gT^A\,]$, 
belonging to the adjoint representation, we have $C=N$, $C_L=C_R 
=(N^2-1)/2N$ which implies $b_n=-1/(N\sqrt{N^2-1})$ and $2b_n/b = 
-1/N^2$.  For example, in the case of $SU(2)$, this ratio $2b_n/b = 
-1/4,$~\cite{aff5} and for $SU(3)$ and $SU(4)$ they are $-1/9$, and 
$-1/16$ respectively.

For the excited state, ${\rm Tr}\;g$, belonging to the singlet 
representation we have $C=0, C_L=C_R= (N^2-1)/2N$.  In this case, $b_n 
= \sqrt{N^2-1}/N$ and the $2b_n/b = 1 - 1/N^2$.  This result is new.  
In case of $SU(2)$, the value $2b_n/b = 3/4$ has been previously 
obtained~\cite{aff5} but for $SU(3)$, $2b_n/b = 8/9$ and for $SU(4)$, 
$2b_n/b = 15/16$ are predictions from our general formula.

\begin{figure}[httb]
\protect \centerline{\epsfxsize=3.0in \epsfbox {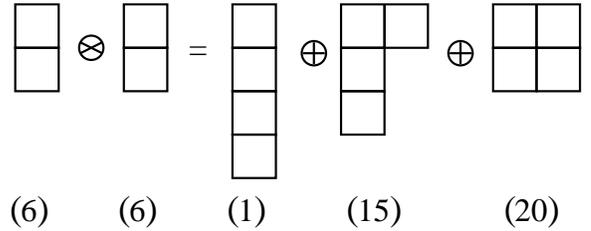}} \vskip 
.6cm \protect \caption{Young tableau for the decomposition of $(6,~6)$ 
representation of $SU(4)$. The number in the parenthesis denotes the 
dimension of the corresponding representation.  }
\protect\label{young}
\end{figure}

We consider one more application of the formula (\ref{bnb}) of current 
interest -- the $SU(4)$ symmetric quantum chain described by the 
$SU(4)_{k=1}$ WZNW model.  In this case, to compute logarithmic 
corrections to the excited states energy we note that there are three 
primary fields with scaling dimensions $\Delta_{p} = 3/4,~ 1,~ 3/4$ 
for $p = 1,~2,~3$ respectively, as seen from Eq.~(\ref{gen}).  The 
case of $p=1$ [and $p=3$], as discussed above, is the fundamental 
field $g$ [and its hermitian conjugate ${\bar g}$] which transforms 
under the $(4,~{\bar 4})$ [and $({\bar 4},~4)$] representation of 
$SU(4)_{L}\times SU(4)_{R}$.  From Eq.~(\ref{xip}), the next lowest energy 
excited states correspond to the primary field operator (denoted by 
$\Psi$) with $\Delta_{2} = 1$.  The field $\Psi$ transforms under the 
$(6,~6)$ representation of $SU(4)_{L}\times SU(4)_{R}$.  This 
$(6,~6)$ representation decomposes as direct sum of a 
singlet, an adjoint and a 20-dimensional representation (as in Fig.  
\ref{young}) under the diagonal $SU(4)$.

We now compute the ratio $2b_n/b$ for the excited states corresponding 
to 20-dimensional representation which has a Young tableau with 2 rows 
and 2 columns. For this representation, the Casimir invariant $C$ in 
Eq.~(\ref{bnb}) is obtained from the formula (\ref{casm}): we find $C=6$, 
and $C_{L}=C_{R}=C_{p=2}=5/2$.  Thus, for the excited states 
corresponding to $\Psi$ in the 20-dimensional subrepresentation of 
$(6,~6)$, we have $2b_n/b = -1/4$.

In summary, we have studied finite size spectrum for one dimensional 
$SU(N)$ symmetric quantum chains using both conformal field theory and 
representation theory of $SU(N)$. We have calculated in general the 
scaling dimensions of all the oscillating modes, and obtained the 
ground state energy as well as correlation lengths of the staggaered 
modes for a finite size system with $SU(N)$ symmetry. Possibilities of 
different types of excited states are also briefly discussed and a 
general formula to compute the logarithmic correction to the excited 
state energies has been derived. The existing results for $N=2,~4$  
agree with the predictions from our general formula.

Authors acknowledge discussions with F. C. Zhang and M. Ma. K. M 
also acknowledge partial support by PRF No. 33611-AC6 and partial
support from Berea College.

\end{document}